\documentclass[11pt]{article}
\usepackage{amsmath}
\usepackage{amsfonts}
\usepackage{amssymb}
\usepackage{graphicx}

\def\be\begin{equation}
 \def\ee{\end{equation}}
\def\bea{\begin{eqnarray}}
\def\eea{\end{eqnarray}}
\def\f{\frac}

\def\l{\label}
\begin{document}
\begin{center}
\LARGE {\bf Standard Cosmological Evolution in f(R) Model to Kaluza
Klein Cosmology }
\end{center}
\begin{center}
{\bf $^a$A. Aghmohammadi\footnote{ agha35484@yahoo.com}}, {\bf
$^{b,c}$Kh. Saaidi}\footnote{ksaaidi@uok.ac.ir},  {\bf $^a$M. R.
Abolhassani\footnote{mrhasani@modares.ac.ir}},\\ {\bf $^b$A.
Vajdi\footnote{Avajdi@uok.ac.ir}}
\\
{ \it $^a$Plasma Physics Research Center, Science and Research
Branch Islamic Azad University of Tehran,
   Iran  }\\
  {\it $^b$Department of Physics, Faculty of Science, University of
Kurdistan, Pasdaran Ave., Sanandaj, Iran}\\
 {\it   $^c$Faculty of Science, Azad University of Sanandaj, Sanandaj, Iran
 }\\
\end{center}
 \vskip 1cm
\begin{center}
{\bf{Abstract}}
\end{center}

In this paper, using $f(R)$ theory of gravity we explicitly
calculate cosmological evolution in the presence of a perfect fluid
source in four- and five-dimensional, space–time in which this
cosmological evolution in self-creation is presented by Reddy {et al
2009 Int. J. Theor. Phys. 48 10}. An exact cosmological model is
presented using a relation between Einstein's gravity field equation
components due to a metric with the same component from $f(R)$
theory of gravity. Some physics and kinematical properties of the
model are also discussed.
\\


\newpage

\section{Introductions}

In recent years, there has been a considerable interest in
alternative theories of gravitation. The observation that universe
appears to be accelerating at present times has caused one of the
greatest problem to modern cosmology. High precision data from
type Ia supernova, the cosmic microwave background and large scale
structure seem to hint that the universe is presently dominated
by an unknown form of energy, dubbed dark energy
\cite{f1,f2,f3,f4,f5,f6}. One obvious contender for the role of
dark energy is Einstein's cosmological constant, but particle
physics failed to predict the correct density. We refer  the
reader to \cite{n7}

Recently, a modification of general  relativity itself was suggested
to explain this accelerating universe \cite{f7, f8, f9, f10, f11,
12, 13, 14, 15, 16, 17}. For review see e.g \cite{n8,n9,n10}. The
assumption is that the Ricci scalar of the Einstein -Hilbert action
is replaced by add a perturbation function $h(R)$ to the Einstein -
Hilbert action\cite{b18}. Recently, it has been claimed that, in all
theories which behave as a power of R at large or small R, standard
cosmological evolution can not be obtained \cite{b19, b20}. These
models have raised much recent interest, due to their perhaps simple
nature. The presence of ghosts and stabilities have been studied
\cite{ n111, n11, n12, n13}.

{ Study of higher -dimensional models are also important because of
the underlying idea that the cosmos at its early stage of evolution
might have had a higher dimensional era \cite{1}. The extra space
reduce to a volume with the passage of time which is beyond the
ability of experimental observation at the moment \cite{1}. Our
motivation to consider the $f(R)$ model in the five dimensional
space times in the presence of a perfect fluid, is that, the same
problem is considered in the self-creation cosmology(SCC), in which
corresponding Mach's Principle(MP) is incorporated in SCC by
assuming the inertial masses of fundamental particles are dependent
upon their interaction with a scalar field $\phi$ coupled to the
large scale distribution of matter in a similar fashion as Brans
Dicke theory (BD). We refer the reader for review see e.g \cite{n14,
1, n15}. However, instead, with recourse to the f (R) model, perhaps
because of their simple nature, and on Q3 modifying part of
geometry, instead of matter in the equation of Einstein, we obtain
the same result. In this paper, we have obtained a cosmological
evolution in the presence of perfect fluid source represents
disordered radiation in four and five dimensional space
time\cite{1}. However, the cosmological evolution  in five
dimensional, in models $f(R)$ in the presence of perfect fluid
source have not been investigated.}
\newpage

\section{Theoretical Framework}
For convenience, we consider a class of modified gravity in which we
add a perturbative function $\epsilon h(R)$, to the Einstein-Hilbert
action, where $\epsilon$ is a small parameter. We consider an action
as
\begin{equation}\l{1}
S=\int\sqrt{-g}\left[ \f{R+\epsilon h(R)}{2}+kL_{m} \right]d^{4}x.
\end{equation}
Here R is a Ricci scalar and $L_m$ is matter Lagrangian. The field
equation, using the metric approach, can be derived from action as,
\begin{eqnarray}
G_{\mu\nu}=-\epsilon\left[G_{\mu\nu}+g_{\mu\nu}\Box
-\nabla_{\mu}\nabla_{\nu}+ {g_{\mu\nu}\over 2}\left(R- {h(R) \over
\varphi(R)}\right)\right]\varphi(R)+kT_{\mu\nu},\l{2}
\end{eqnarray}
where $G_{\mu\nu}$ and  $T_{\mu\nu}$ are Einstein and stress-energy
tensors respectively, $\Box \equiv \nabla_{\alpha}\nabla^{\alpha} $
and $\varphi(R)=dh(R)/dR$.  Although this theory can be written as
scaler tensor theory \cite{13,b21}, we shall not use the conformal
transformation, we complete all our calculations in the Jordan
frame, given by the action above. We consider the metric given by
\begin{equation}\l{3}
ds^2=dt^2-a^2{(t)}(dr^2+r^2d\Omega^2).
\end{equation}
 In which the space-time is assumed to
be of flat Friedmann -Robertson-Walker (FRW), the components of
 Ricci tensor can be written in terms of the scale factor, $a(t)$,  as
\begin{eqnarray}
R^t_t&=&3\f{\ddot{a}}{a},\l{4} \\
R^r_r&=&R^{\theta}_{\theta}=R^{\phi}_{\phi}=\f{a\ddot{a}+2\dot{a}^2}{a^2},\l{5}
\end{eqnarray}
where the over dot denotes differentiation  with respect to  cosmic
time. The components of Einstein tensor in terms  of the scale
factor, $a(t)$, in the flat Friedmann-Robertson-Walker line element,
can be find as following
\begin{eqnarray}
G^t_t&=& 3\left(\f{\dot{a}}{a}\right)^2
,\l{6}\\
G^r_r&=&G^{\theta}_{\theta}=G^{\phi}_{\phi}=-(\f{2a\ddot{a}+\dot{a}^2}{a^2}).\l{7}
\end{eqnarray}
The components of d'Alembertian in the flat
Friedmann-Robertson-Walker line element are as following
\begin{eqnarray}
\nabla_t\nabla^t&=&\f{{\partial}^2}{{\partial t}^2},\l{8}\\
\Box&=&\f{{\partial}^2}{{\partial t}^2}+\f{3\dot{a}}{a}\f{\partial}{\partial t}, \l{9}\\
\nabla_r\nabla^r=\nabla_{\theta}\nabla^{\theta}&=&\nabla_{\phi}\nabla^{\phi}=\f{\dot{a}}{a}\f{\partial}{\partial
t }.\l{10}
\end{eqnarray}

The energy-momentum tensor, $T_{\mu\nu}$, for a perfect fluid
distribution is given by
\begin{equation}\l{11}
T_{\mu\nu}=\left(p+\rho\right)u_{\mu}u_{\nu}-pg_{\mu\nu}.
\end{equation}
Here, $p$ is the isotropic pressure, $\rho$ the energy density and
$u_{\mu}$ represents the four velocity of the fluid. Corresponding
to the line element given by (\ref{3}) the four velocity vector
$u_{\mu}$ satisfies the equation
\begin{equation}\l{12}
g_{\mu\nu}u^{\mu}u^{\nu}=1.
\end{equation}
In a co-moving coordinate system the components  of Einstein tensor
with the help of field equation, using the metric approach,  can be
derived from action (\ref{1}). In the field equation (\ref{2}), if
$h(R)/\varphi(R)=0$ as $R\rightarrow 0$ i.e.;
\begin{equation}\l{13}
\lim_{R\rightarrow 0}\f{h(R)}{\varphi(R)}\rightarrow0,
 \end{equation}
  we can neglect  $h(R)/\varphi(R)$, and then
  we have
\begin{eqnarray}
 G^{\nu}_{\mu}&=&-\epsilon\left(\delta^{\nu}_{\mu}\Box-\nabla^{\nu}\nabla_{\mu}\right)\varphi(R)+T^{\nu}_{\mu},\l{14}\\
 G^t_t&=&-\epsilon\left( 3\f{\dot{a}}{a}\f{\partial}{\partial t}\right)\varphi+k\rho ,\l{15}\\
  G^r_r=G^{\theta}_{\theta}=G^{\phi}_{\phi}&=&-\epsilon\left(\f{{\partial}^2}{{\partial t}^2}+2\f{\dot{a}}{a}\f{\partial}{\partial t}
 \right)\varphi(R)+kp .\l{16}
 \end{eqnarray}
Where $\rho$ and $p$ are components of $T^{\nu}_{\mu}$.
Furthermore, from
 contracting field equation (\ref{14}), we can find
\begin{equation}\l{17}
R=3\epsilon\Box\varphi,
\end{equation}
from combination equations (\ref{6},\ref{7}) with
(\ref{15},\ref{16}), one obtains

\begin{eqnarray}
3\left(\f{\dot{a}}{a}\right)^2&=&-\epsilon\left(\f{3\dot{a}}{a}\f{\partial}{\partial
t}\right)\varphi+k\rho ,\l{18}\\
\f{2a\ddot{a}+\dot{a}^2}{a^2}&=&\epsilon\left(\f{\partial^2}{{\partial
t}^2}+2\f{\dot{a}}{a}\f{\partial}{\partial t} \right)\varphi-kp
.\l{19}
\end{eqnarray}
The field equations (\ref{17},\ref{18},\ref{19}) are three
independent equation in four unknown variable $ a, \varphi, \rho,
p$. Hence to get a determinate solution and correspondence with the
standard Einstein cosmology solution one has to assume a physical or
mathematical condition. We choose  the scale factor as
\begin{equation}\l{20}
a(t)=t^{\gamma},
\end{equation}
and the equation  of state as
\begin{equation}\l{21}
\rho=3p,
\end{equation}
which represents disordered radiation in four dimensional space.
Now, using equations (\ref{20},\ref{21}) the field equations
(\ref{17},\ref{18},\ref{19}) yields an exact solution for $
a,\varphi,\rho,p$. According to equations (\ref{18}, \ref{19}) and
(\ref{20}), we  can choose  $\varphi$ as :
\begin{equation}\l{22}
\varphi=\alpha\ln t +c,
\end{equation}
where $\alpha$ and $c$ are  arbitrary constants.  Substituting
equations (\ref{22}, \ref{18}, \ref{19}), into equation (\ref{17}),
power-law solution can be identified as
\begin{equation}\l{23}
\gamma=\f{1}{2}+\f{\alpha\epsilon}{4}.
\end{equation}
By substituting equation  (\ref{23}) into equation (
\ref{18},\ref{21}), we can arrive at
\begin{eqnarray}
\rho&=&\f{3}{4kt^2}\left(1+\alpha\epsilon\right),\l{24}\\
p&=&\f{1}{4kt^2}\left(1+\alpha\epsilon\right),\l{25}\\
H&=&\f{2+\alpha\epsilon}{4t},\hspace{1cm}\l{26}\\ q&>&0 \l{27},
\end{eqnarray}
which with $\epsilon=0$, solutions above are correspond with epochs
of radiation domination. The energy density $\rho$, the isotropic
pressure $p$, tend to zero as time increases indefinitely. Also
Hubble's parameter, $H$, tend to zero as $t\rightarrow \infty$. The
positive value of the deceleration parameter, $q$, shows that the
model decelerates in the standard way.

\section{Kaluza-Klein Cosmological in f(R) Theory of Gravity  }
Once more we consider action (\ref{1}) together with field equation
(\ref{2}) that yields to variation of action (\ref{1}) for the line
element of five dimensional 5D Kaluza-Klein space time given by
\begin{equation}\l{28}
ds^2=dt^2-a^2{(t)}\left({dx}^2+{dy}^2+{dz}^2\right)-A^2{(t)}{d\psi}^2.
\end{equation}
In Kaluza-Klein space time, the components of   Ricci tensor and
Ricci scaler can be written in terms of the scale factor, $a(t)$ and
$A(t)$. As a result, we find
\begin{eqnarray}
R_{tt}&=&\f{3\ddot{a}}{a}+\f{\ddot{A}}{A},\l{29}\\
R_{{\psi}{\psi}}&=& -\f{3A{\dot{a}}\dot{A}}{a}-A\ddot{A},\l{30}\\
R_{xx}=R_{yy}&=&R_{zz}=-2\left(\dot{a}\right)^2-\f{a{\dot{a}}\dot{A}}{A}-a\ddot{a},\l{31}\\
R&=&6\left(\f{{{\dot{a}}}^2}{a^2}+\f{\ddot{a}}{a}\right)+6\f{\dot{a}\dot{A}}{aA}+2\f{\ddot{A}}{A}
.\l{32}
\end{eqnarray}
The components  of Einstein tensor  in terms of the scale factors
$a(t)$ and  $A(t)$, in the Kaluza Klein space-time line element, can
be found as following :
\begin{eqnarray}
G^t_t&=&3\left(\f{\dot{a}}{a}\right)^2+3\f{\dot{a}\dot{A}}{aA},\l{33}\\
G^{\psi}_{\psi}&=&-3\left(\f{\dot{a}}{a}\right)^2-3\f{\ddot{a}}{a},\l{34}\\
G^x_x&=&-\left(\f{\dot{a}}{a}\right)^2-2\f{\ddot{a}}{a}-2\f{\dot{a}\dot{A}}{aA}-\f{\ddot{A}}{A}
.\l{35}
\end{eqnarray}
Dolambrian and its components to the Kaluza-Klein line element are
as following :
\begin{eqnarray}
\nabla_t\nabla^t&=&\f{{\partial}^2}{{\partial t}^2},\l{36}\\
\nabla_{\psi}\nabla^{\psi}&=&\f{\dot{A}}{A}\f{\partial}{\partial t},\l{37}\\
\nabla_x\nabla^x=\nabla_y\nabla^y&=&\nabla_z\nabla^z=\f{\dot{a}}{a}\f{\partial}{\partial
t},\l{38}\\
\Box&=&\f{{\partial}^2}{{\partial
t}^2}+\left(\f{3\dot{a}}{a}+\f{\dot{A}}{A}\right)\f{\partial}{\partial
t}.\l{39}
\end{eqnarray}
In the Kaluza-Klein space time and a co-moving coordinate system the
components of Einstein tensor with the help of field equations
(\ref{14}, \ref{15}, \ref{16}) are given from variation of action
(\ref{1}) as following

\begin{eqnarray}
G^t_t&=&-\epsilon\left(\f{3\dot{a}}{a}+\f{\dot{A}}{A}\right)\f{\partial}{\partial
t}\varphi(R)+k\rho ,\l{40}\\
G^{\psi}_{\psi}&=&-\epsilon\left(\f{{\partial}^2}{{\partial
t}^2}+\f{3\dot{a}}{a}\f{\partial}{\partial t}\right)\varphi(R)+kp ,\l{41}\\
 G^x_x=G^y_y=G^z_z&=&-\epsilon\left(\f{{\partial}^2}{{\partial
 t}^2}+\left(\f{2\dot{a}}{a}+\f{\dot{A}}{A}\right)\f{\partial}{\partial
 t}\right)\varphi(R)+kp .\l{42}
\end{eqnarray}
Where $\rho$ and $p$ are component of the $T^{\mu}_{\nu}$.
Furthermore, from contraction field equation (\ref{14}), we can
obtain as:
\begin{equation}\l{43}
R=\f{8}{3}\epsilon\left(\f{{\partial}^2}{{\partial
t}^2}+\left(\f{3\dot{a}}{a}+\f{\dot{A}}{A}
\right)\f{\partial}{\partial t} \right)\varphi.
\end{equation}
From equations (\ref{29}-\ref{39}) we obtains:

\begin{eqnarray}
3\left(\f{{\dot{a}}^2}{a^2}+\f{\dot{a}\dot{A}}{aA}
\right)&=&-\epsilon\left(\f{3\dot{a}}{a}+\f{\dot{A}}{A}\right)\f{\partial}{\partial
t}\varphi(R)+k\rho,\l{44}\\
\f{{\dot{a}}^2}{a^2}+2\f{\ddot{a}}{a}+2\f{\dot{a}\dot{A}}{aA}+\f{\ddot{A}}{A}&=&\epsilon\left(\f{{\partial}^2}{{\partial
t}^2}+\left(
\f{2\dot{a}}{a}+\f{\dot{A}}{A}\right)\f{\partial}{\partial t}
\right)\varphi(R)-kp ,\l{45}\\
 +3\f{{\dot{a}}^2}{a^2}
+3\f{\ddot{a}}{a}&=&\epsilon\left(\f{{\partial}^2}{{\partial
t}^2}+\f{3\dot{a}}{a}\f{\partial}{\partial t} \right)\varphi(R)-kp.
\l{46}
\end{eqnarray}
The set field equations (\ref{43}-\ref{46}) are four independent
equation in five unknowns $\rho, p, a, A, \varphi$. Hence to get a
determinate solution one has to assume a physical or mathematical
condition. We solve the above set of field equations with the
equation of state trace $T=0$
\begin{equation}\l{47}
\rho=4p,
\end{equation}
which is analogous to the equation of state before $\rho=3p$ which
represents disordered radiation in four dimensional space. To obtain
a determinate solution we also use a relation between the metric
potentials as
\begin{equation}\l{48}
A=\xi a^n.
\end{equation}
Where $\xi ,n$ are constants. Substituting  equations (\ref{20},
\ref{47}, \ref{48}) into (\ref{44}, \ref{45}, \ref{46}),we can
obtain $\gamma$, and the other unknowns parameters  for a small
$\epsilon$, as following
\begin{eqnarray}
\gamma&=&\f{3+n}{\eta}\left(1+\f{2}{3}\zeta\alpha\epsilon+ \ldots\right),\l{49}\\
A&=&\xi
t^{n\f{3+n}{\eta}\left(1+\f{2}{3}\zeta\alpha\epsilon +\ldots\right)},\l{50}\\
H&=&\f{1}{t}\f{3+n}{\eta}\left(1+\f{2}{3}\zeta\alpha\epsilon + \ldots\right),\l{51}\\
q&=&-\f{\left(\gamma-1\right)}{\gamma},\l{52}\\
\rho&=&\f{(3+n)^2}{kt^2\eta}\left(\f{3(n+1)}{\eta}+(1+\f{4(n+1)\zeta}
{\eta})\alpha\epsilon + \ldots  \right),\l{53}\\
p&=&\f{(3+n)^2}{4kt^2\eta}\left(\f{3(n+1)}{\eta}+(1+\f{4(n+1)\zeta}
{\eta})\alpha\epsilon + \ldots  \right),\l{54}
\end{eqnarray}
 where
\begin{eqnarray}
\eta&=& n^2+3n + 6, \\
\zeta&=& 1 - \f{n^2 +1}{(n+3)^2}.
\end{eqnarray}
We require that the model decelerates in the standard way, so that
$q$ must be bigger than zero. Therefore, we find an extra condition
on $\gamma$ as
\begin{equation}\l{55}
0 <\gamma <1,
\end{equation}
 and then
\begin{equation}\l{55}
n\neq -3.
\end{equation}
For  considering  the  stability of  this model of $f(R)$, we must
obtain the second derivation of $f(R)$ with respect to R. It is well
known that, for an arbitrary $f(R)$ model, if ${d^2f(R) \over dR^2 }
> 0 $, then that model is stable \cite{ n111, n13}. Therefore,
using equations (\ref{22}, \ref{43}), we can obtain
\begin{equation}\l{56}
{d^2f(R) \over dR^2}=
\epsilon\f{d\varphi}{dR}=\f{3t^2}{16\epsilon((n+3)\gamma -1)}.
\end{equation}
Hence, the requirement for stability of our model, is that for
$\epsilon>0$, $n$ must satisfy
\begin{equation}\l{57}
 n>\f{1}{\gamma}-3,
\end{equation}
and for the case $\epsilon<0$, $n$ must satisfy
\begin{equation}\l{58}
    n< \f{1}{\gamma}-3
\end{equation}
 Also, for considering the singularity   of this  $f(R)$ model of
 gravity
    at initial time, we need  the Ricci scalar as a function of time.
 Therefore, with combining  equations, (\ref{20}, \ref{22},\ref{43}, \ref{48}) and (\ref{49}),
 one can obtain the Ricci scalar as following
\begin{equation}\l{59}
R =\f{8}{3}\epsilon\alpha\left(-1+(3+n)\gamma\right)\f{1}{t^2}.
\end{equation}
It is clearly seen that, $ R = \infty $ as $t$ tends to zero,
i.e.;
\begin{equation}
\lim_{t\rightarrow 0}R \rightarrow \infty,
\end{equation}
 this
means that, this model has initial singularity.
 Thus, the five dimensional Kaluza-Klein   cosmological line
element corresponding to the above solutions can be written in
$\epsilon=0$ by:
\begin{equation}\l{60}
{ds}^2={dt}^2-\left(t^{\f{2n+6}{6+3n+n^2}}\right)\left({dx}^2+{dy}^2+{dz}^2\right)-{\xi}^2t^{n\left(\f{2n+6}{6+3n+n^2}\right)}{d\phi}^2
\end{equation}
\section{Some Physical Properties Of the Model}
 Equation (\ref{49}$\cdots$\ref{54})
represents an exact five dimensional Kaluza-Klein cosmological in
the framework of $f(R)$ theory of gravity in the presence of perfect
fluid source. We observe  that the $f(R)$ theory of gravity has
initial singularity, (\ref{59}), and  we have seen  that with $n$
defined by equations (\ref{57},\ref{58}), the requirement of
stability is satisfied. For the Eq(\ref{60}), the physical and
kinematical variables which are important in cosmology are
$\rho,p,q,H$ the energy density  $\rho$, the isotropic pressure p,
tend to zero as time increases indefinitely. For this model Hubble
parameter tend to zero as $t\rightarrow \infty$, the positive value
of the deceleration parameter q, shows that to the both four and
five dimension decelerates in the standard way.

\section{Discussion}

A modification of gravity has been suggested in the form of
$R+\epsilon h(R)$ theories, in the presence of perfect fluid
source. The five dimensional Kaluza-Klein and four dimensional
line element in the $f(R)$ theory of gravity  has stability and
initial singularity. The model is expanding, non-rotating and
decelerates in the standard way.

\end{document}